\pdfoutput=1

\documentclass[aps,pra, amsmath,onecolumn, notitlepage, 12pt]{revtex4-1}
\usepackage{graphicx}
\usepackage{bm}
\usepackage{titlesec}
\usepackage{setspace}

\def\beq{\begin{equation}}
\def\eeq{\end{equation}}
\def\bea{\begin{eqnarray}}
\def\eea{\end{eqnarray}}

\begin{document}
\setstretch{1.05}

\title[Enhanced Conductivity Along Lateral Homojunction Interfaces]{Enhanced Conductivity Along Lateral Homojunction Interfaces of Atomically Thin Semiconductors}
\author{Ying Jia, Teodor K. Stanev, Erik J. Lenferink}
\author{Nathaniel P. Stern}
\email{n-stern@northwestern.edu}
\affiliation{Department of Physics and Astronomy, Northwestern University, Evanston, Illinois 60208 USA\vspace{2em}}

\begin{abstract}
\vspace{1em}
Energy band realignment at the interfaces between materials in heterostructures can give rise to unique electronic characteristics and non-trivial low-dimensional charge states. In a homojunction of monolayer and multilayer MoS$_2$, the thickness-dependent band structure implies the possibility of band realignment and a new interface charge state with properties distinct from the isolated layers. In this report, we probe the interface charge state using scanning photocurrent microscopy and gate-dependent transport with source-drain bias applied along the interface. Enhanced photoresponse observed at the interface is attributed to band bending. The effective conductivity of a material with a monolayer-multilayer interface of MoS$_2$ is demonstrated to be higher than that of independent monolayers or multilayers of MoS$_2$. A classic heterostructure model is constructed to interpret the electrical properties at the interface. Our work reveals that the band engineering at the transition metal dichalcogenides monolayer/multilayer interfaces can enhance the longitudinal conductance and field-effect mobility of the composite monolayer and multilayer devices.

\end{abstract}

\maketitle

\section{Introduction\vspace{-.25em}}
Transition metal dichalcogenides (TMDCs) are layered crystals with thickness-dependent band gaps, 1.29~eV in a multilayer (ML) and 1.88~eV for a monolayer (1L) of MoS$_2$~\cite{bandstructure} for example. The distinct electronic structure implies an energy band discontinuity at the interface in 1L and ML MoS$_2$. Recent Kelvin probe force microscopy has detected different electron affinities in 1L and ML MoS$_2$ and the conduction band offset at the boundary between two regions of the same region material but with different layer numbers~\cite{KelvinP} (which we refer to here as a 1L/ML homojunction).  According to the electron affinity model, band realignment should occur near the interface and may create unique interfacial electronic states with properties distinct from those of the parent compounds~\cite{semiconductorBook1}. Examples of such emergent interfacial properties in heterostructures are high-mobility two-dimensional electron gases in Al$_x$Ga$_{1-x}$As/GaAs~\cite{2DEG}, and interface superconductivity in Bi$_2$Te$_3$/FeTe~\cite{interfaceSC}. Distinct from the `conventional’ semiconductor heterojunctions, experimental and calculated results have suggested localized metallic states at the 1D edges of 1L and few-layer MoS$_2$~
\cite{Wu2016, edge17, edge18, edge15, edge16, Li2008, Xiao2015}. In a lateral 1L/ML MoS$_2$ homojunction (Fig. 1a), such an edge state of ML MoS$_2$ could influence the band realignment at the interface, complicating the electronic structure and the properties of the interface electrons.

Recent studies on TMDC 1L/ML homojunctions were primarily focused on photocurrent generation at the interfaces with source-drain contacts on opposite sides of the 1L/ML junctions~\cite{KelvinP, H, e}. Band realignment has been confirmed though the details are still under debate~\cite{e}. Current versus voltage ($I – V$) measurements were also performed with currents applied across the junction~\cite{H, e} with non-linear $I-V$ curves consistently observed. The existence of conducting interfacial edge charge states at multilayer junctions separating distinct TMDC thicknesses has been measured~\cite{Wu2016}, but the consequences of layer-sensitive composite structures for low-dimensional interfacial transport at the monolayer limit have not yet been investigated. In particular, the longitudinal transport properties of 1L/ML interfaces remain unstudied.

Here, we fabricate lateral MoS$_2$ 1L/ML homojunctions and investigate the longitudinal transport properties of the interface charge state with source-drain bias applied along the interface. Interfacial band bending is confirmed by enhanced photoresponse along the junction boundary using scanning photocurrent microscopy. Electric conductivity of the 1L and ML composite devices is measured and compared with that of the independent 1L and ML MoS$_2$ of the same flake. The higher effective conductivity of the composite devices indicates a notable contribution from the interface charge state. Our results outline an experimental approach to studying interfacial conduction properties of electronic states at layered TMDC homojunctions.

\section{Methods\vspace{-.25em}}
The MoS$_2$ flakes were mechanically exfoliated from undoped crystals and deposited on heavily doped silicon substrates covered with 285-nm-thick SiO$_2$ using elastic-film-assisted micro-mechanical exfoliation~\cite{gel}. MoS$_2$ flakes with both 1L and ML regions were identified using optical microscopy and then confirmed using atomic force microscopy. Fig. 1b shows a typical as-exfoliated sample. This sample consists of 8 layers on one side and 1 layer on the other side, as shown in the inset of Fig. 1b. The 1L/ML structures were patterned into individual pieces with Au electrodes using e-beam lithography and SF$_6$ reactive ion etching~\cite{SF6}. Each of the 1L, ML, and 1L$+$ML devices were independently measured with a 4-point method. All devices were operated in a back-gating configuration to maintain the direct accessibility of an excitation laser with the heavily doped Si substrates acting as the back gates.

\begin{figure}
\centering
\includegraphics[width=8.5cm]{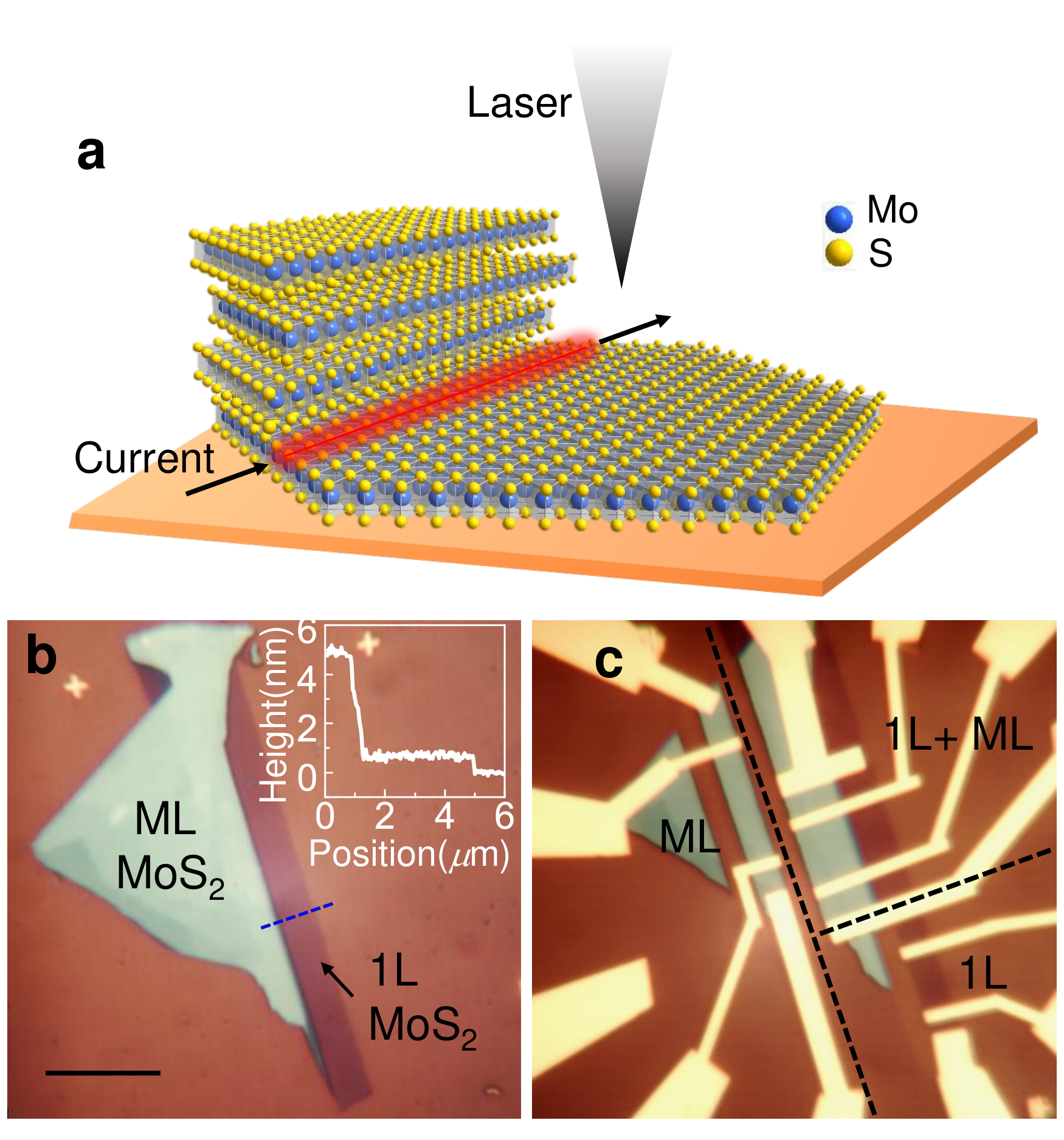}
\caption{{Schematic and optical image of a MoS$_2$ homojunction.} (a) Schematic illustration of the 1L/ML homojunction and the experiment geometry. (b) Optical image of an as-exfoliated 1L/ML flake. The inset is the thickness profile along the blue dotted line. The scale bar is 10~$\mu$m. (c) Optical image of the 1L, ML, and 1L/ML devices fabricated from a single flake.
} \label{figure1}
\end{figure}

\section{Results\vspace{-.25em}}
Scanning photocurrent microscopy was used initially to probe the generation of carriers and band bending at the 1L/ML homo-interface. The measurement was performed in high vacuum (less than 1~mTorr) and at room temperature with laser excitation power of 70 $-$ 80~$\mu$W at several wavelengths as noted. In few-layer MoS$_2$, the primary photocurrent mechanisms involve two processes, exciton generation by a photon with energy higher than optical band gap and free carrier creation by electric-field-assisted dissociation~\cite{Ipcgeneration}. At the 1L/ML interface, band bending induces a high local built-in electric field which can dissociate excitons efficiently~\cite{suspend} and enhance local photoresponse.  With our contact geometry, the built-in field is perpendicular to the current channel, and cannot drive the dissociated free carriers to the contacts. The magnitude of photoresponse is expected to be lower than that with source-drain contacts on opposite sides of the interface. A source-drain bias is required to produce photocurrent in the circuit.

\begin{figure}
\centering
\includegraphics[width=8.5cm]{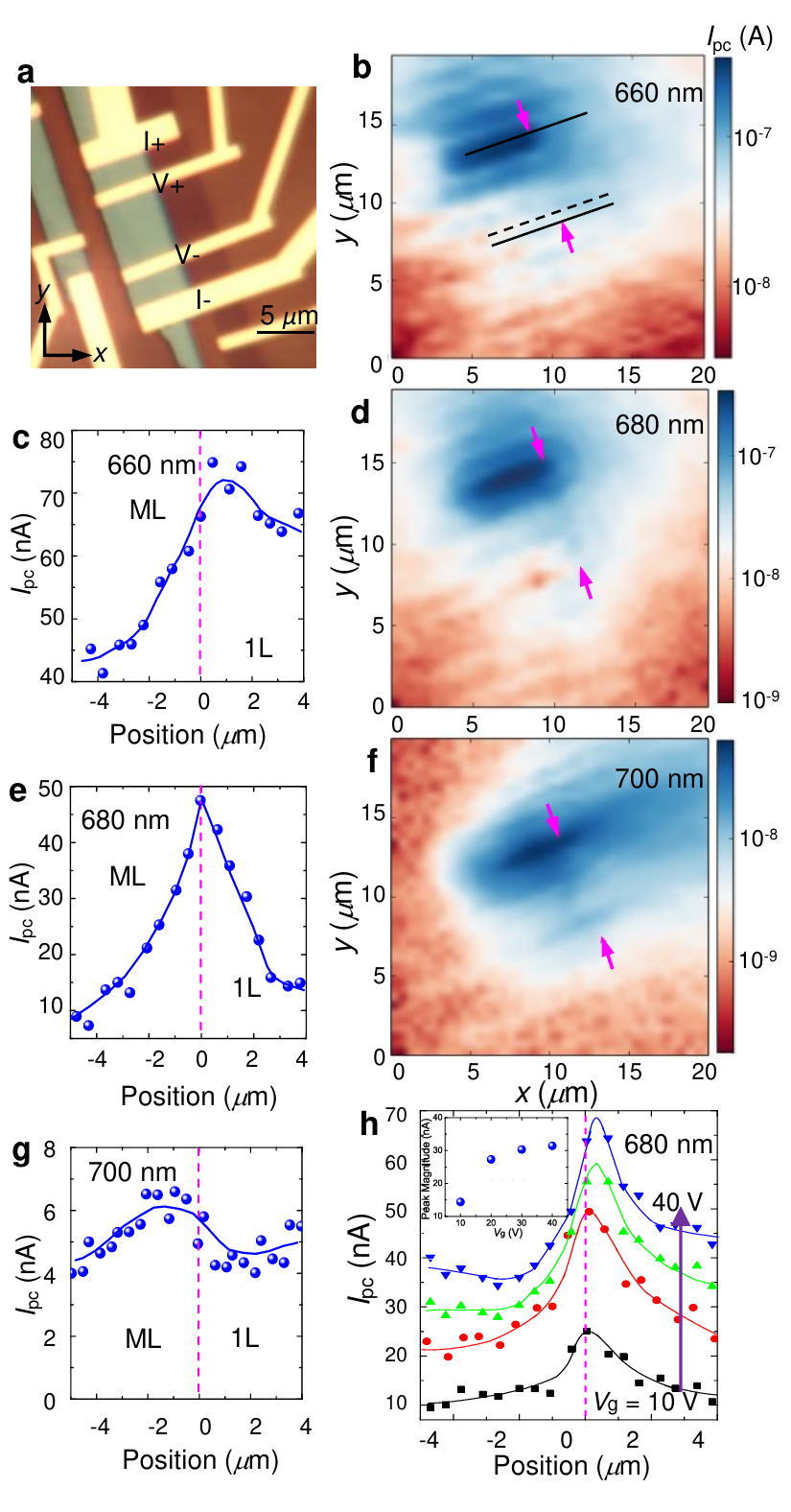}
\caption{{Photocurrent generation in 1L/ML homostructures at various excitation wavelengths.} (a) Optical image of the 1L$+$ML device. A bias voltage is applied between the $V_+$ and $V_-$ contacts. (b-f) Photocurrent intensity spatial maps with line profiles along the dashed line in b at excitation wavelength of 660~nm (b-c), 680~nm (d-e), and 700~nm (f-g). $I_{\rm pc}$ is plotted on a log scale in spatial maps and with normal scale in line profiles. The black solid lines in b indicate inner edges of the $V_+$ and $V_-$ contacts. The pink arrows mark the 1L/ML boundary. In c, e, and g, the pink dashed lines separate the 1L and ML regions. The solid lines are guides for eyes. The log scale reduces the prominence of the contact region, allowing the edge signal to be seen in the map, although the edge features are clearest in the line profiles. (h) Profiles of photocurrent along a fixed scan line perpendicular to the interface at $V_{\rm g}$ = 10, 20, 30, and 40~V. The wavelength of the excitation laser is 680~nm. Successive curves are offset by 7~nA for better visualization. The inset shows the peak magnitude as a function of the gate voltage.} \label{figure2}
\end{figure}

Fig. 2a shows the optical image of the scanned area (20 $\times$ 20~$\mu$m$^2$). The widths of the 1L and ML channels are 5~$\mu$m and 4~$\mu$m, respectively. A bias voltage of 0.2~V was applied between the $V_+$ and $V_-$ contacts. The photocurrent maps were acquired with excitation wavelengths 660~nm~($h\nu$ = 1.88~eV, Fig. 2b), 680~nm~($h\nu$ = 1.83~eV, Fig. 2d), and 700~nm~($h\nu$ = 1.77~eV, Fig. 2f). A back-gate voltage of 20 V was applied during the scan. In our experiment, the intensity of the reflected laser and the photocurrent ($I_{\rm pc}$) generated in the device were simultaneously recorded at each position, allowing the spatial photocurrent map to be correlated with the device geometry. The inner edges of the $V_+$ and $V_-$ contacts and the 1L/ML boundary in Fig. 2a are marked by black solid lines and arrows in Fig. 2b. The highest $I_{\rm pc}$ intensity is observed at the inner edge of the $V_+$ contact, which can be attributed to the local electrical field at the Au/MoS$_2$ interface and the bias voltage applied between the $V_+$ and $V_-$ contacts~\cite{Ipcgeneration}. To visualize the spatial dependence of $I_{\rm pc}$ in other areas, $I_{\rm pc}$ is plotted with a log scale in Fig. 2b, 2d, and 2f. The photocurrent line profiles at various wavelengths were also recorded across the 1L/ML boundary, shown in Fig. 2c, 2e, and 2g. To avoid the influence of the high-intensity photocurrent at the $V_+$ contact, the line is chosen close to the $V_-$ contact, as shown by the black dashed line in Fig. 2b.

Under illumination of $\lambda$ = 680~nm, a region with enhanced photocurrent is observed between the two arrows in Fig. 2d, overlapping with the 1L/ML boundary. A $I_{\rm pc}$ peak located at the boundary is also observed in the line profile in Fig. 2c. This confirms that the bent bands at the 1L/ML interface can increase local photocurrent generation. The enhanced photocurrent at the 1L/ML boundary is observed for various gating voltages $V_{\rm g}$, ranging from 10~V to 40~V (Fig. 2h). The $V_{\rm g}$ dependence of peak magnitude is much weaker than that observed on MoS$_2$/metal boundaries~\cite{PCmetal} because the global shifting of the back gate couldn't change the built-in potential at the 1L/ML interface. This observation is consistent with the $V_{\rm g}$-dependent $I_{\rm pc}$ with contacts on opposite sides of the junction~\cite{KelvinP, H}. At $\lambda$ = 660~nm, the photon energy very nearly matches the optical band gaps in 1L and ML MoS$_2$. Photocurrent arising from other effects such as hot carriers~\cite{hotcarrier} and photo-thermoelectric effect~\cite{thermoelectric} increases in the interior of 1L and ML~\cite{discovery, phototransistor} and reduces  the intensity contrast at the 1L/ML boundary. At $\lambda$ = 700~nm, the photon energy is too low to effectively create excitons in MoS$_2$. The overall photocurrent intensity is strongly reduced and no enhancement can be observed at the 1L/ML boundary.

Here, the photocurrent enhancement at the 1L/ML boundary is interpreted as arising from bent bands, but the possibility of a new state with optical band gap 1.83~eV cannot be ruled out. Unusual edge states of TMDCs have been observed with a band gap smaller than that of the interior~\cite{edge1, edge2}. An optically active band gap of 1.42~eV was theoretically predicted and experimentally observed in MoS$_2$/WS$_2$ bilayer vertical heterostructures~\cite{CVD3ref25, CVD3ref26, CVD3}. Nevertheless, the enhanced photocurrent suggests that the 1L/ML interface exhibits new properties associated with the unique band structure of this boundary.

To probe the conducting properties of the interface charge state, we measured transport of devices with current flowing along a 1L and ML boundary (1L$+$ML devices). The conductance of multiple composite 1L$+$ML devices were measured and compared with the independent 1L and ML devices fabricated from the same sample. For each set of the 1L, ML, and 1L$+$ML devices, the devices originate from the same single-crystalline flake and were processed with the same procedures. It can thus be assumed that the conductivity of the 1L (or ML) device equals to that of the 1L (or ML) channel in the 1L$+$ML device. This control design allows us to compare the effective conductivity of the 1L$+$ML device $\sigma_{\rm eff}=G\cdot\frac{L}{W_{\rm m}+W_{1}}=\frac{I_{\rm sd}}{V_{\rm sd}}\cdot\frac{L}{W_{\rm m}+W_{1}}$ ($G$ is the net conductance of the 1L$+$ML device, $W_{\rm m}$, $W_{1}$, and $L$ are the widths and length of the ML and 1L channels in the 1L$+$ML device) with the conductivity of the independently measured 1L and the ML devices, $\sigma_1$ and $\sigma_{\rm m}$, respectively. Typically $\sigma_{\rm m}$ is higher than $\sigma_1$~\cite{SonglinLi}. If the conductivity of the interface conduction channel is close to $\sigma_1$ or $\sigma_{\rm m}$, a naive analysis $\sigma_{\rm m}>\sigma_{\rm eff}>\sigma_1$ should be observed. However, in all our measured devices, it is found that $\sigma_{\rm eff}>\sigma_{\rm m}>\sigma_1$, suggesting an enhanced conductivity at the 1L/ML interface. Two examples are shown in Fig. 3.

Fig. 3a shows the room-temperature gate-dependent conduction properties of the 1L, ML, and 1L$+$ML devices from 4-point measurement in Fig. 1c. Linear $I_{\rm sd}-V_{\rm sd}$ curves are observed in all the devices confirming Ohmic contacts (the inset of Fig. 3a and Fig. S3 in the Supplementary Information). The gating characteristics are plotted as an effective conductivity $G\frac{L}{W}$. The effective conductivity $\sigma_{\rm eff}$ defined in this manner for the composite 1L$+$ML device represents the conductivity assuming the entire device was uniform; it is not the actual conductivity of any conduction channel in the device. The effective conductivity $G\frac{L}{W}$ is equivalent to the actual conductivity $\sigma_1$ and $\sigma_{\rm m}$ for the 1L and the ML devices, respectively. Of the three devices, the 1L$+$ML has the highest effective conductivity over the full range of $V_{\rm g}$, indicating that the interface is more conductive than the 1L or ML regions. The effective field-effect mobility of the 1L$+$ML device can be calculated using $\mu_{\rm FE-eff}=\frac{1}{C_{\rm i}}\frac {\partial\sigma_{\rm eff}}{\partial V_{\rm g}} $ and compared with the field-effect mobilities of the 1L and ML devices, $\mu_{\rm FE-1}=\frac{1}{C_{\rm i}}\frac {\partial\sigma_1}{\partial V_{\rm g}}$ and $\mu_{\rm FE-m}=\frac{1}{C_{\rm i}}\frac {\partial\sigma_{\rm m}}{\partial V_{\rm g}}$, where $C_{\rm i}= 1.3 \times 10^{-4}$~F/m$^{2}$ is the capacitance per unit area of our 285-nm-thick SiO$_2$ layer~\cite{electronics}. The effective mobilities of these devices are $\mu_{\rm FE-1}=$4.1~cm$^{2}$/Vs, $\mu_{\rm FE-m}=$11.4~cm$^{2}$/Vs, and $\mu_{\rm FE-eff}=$20.2~cm$^{2}$/Vs. The value of $\mu_{\rm FE-1}$ is within the range of $0.1-10$~cm$^{2}$/Vs expected for uncapped, back-gated monolayer MoS$_2$ FETs on SiO$_2$/Si substrates. The increased conductivity and mobility of ML MoS$_2$ relative to the 1L device is consistent with previous reports~\cite{SonglinLi}. As with the effective conductivity $\sigma_{\rm eff}$, the effective field-effect mobility $\mu_{\rm FE-eff}$ of the 1L$+$ML device is higher than that of the 1L and ML channel, which further confirms that the 1L/ML interface has a significant impact on the conduction properties of the composite device.

The observed relative ordering of $\sigma_{\rm eff}>\sigma_{\rm m}>\sigma_1$ and $\mu_{\rm FE-eff}>\mu_{\rm FE-m}>\mu_{\rm FE-1}$ is reproduced in additional 1L$+$ML devices. Fig. 3b compares the conductivity (or effective conductivity) of a 1L, a 2L, and a 1L$+$2L two-terminal FETs again fabricated from a same MoS$_2$ flake. It is clearly observed that the effective conductivity of the 1L$+$2L device is higher than the conductivity of the 1L and the 2L devices. The field-effect mobility (or effective mobility) values are $\mu_{\rm FE-1}=$2.8~cm$^{2}$/Vs, $\mu_{\rm FE-m}=$11.0~cm$^{2}$/Vs, and $\mu_{\rm FE-eff}=$15.2~cm$^2$/Vs. Transport properties of an additional 1L$+$ML device, temperature dependence, and a summary of all measured devices are presented in the Supplementary Information. As shown in the Supplementary, the enhanced effective conductivity persists at low temperatures and with different gate voltages. In every measured 1L$+$ML device, the presence of the 1L/ML interface enhances the effective conductivity over that of the independent 1L and ML channels.

\begin{figure}
\centering
\includegraphics[width=8.5cm]{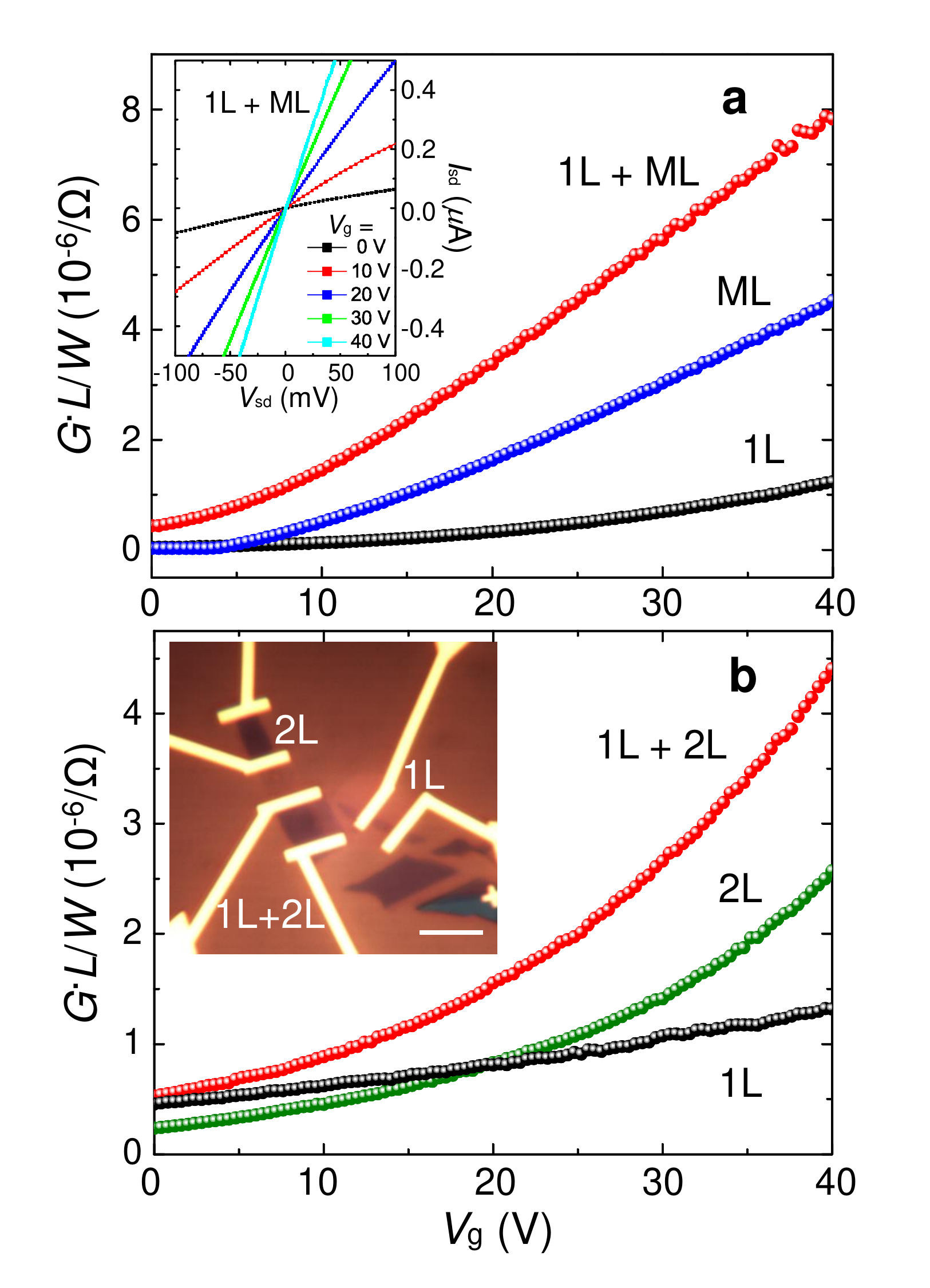}
\caption{{Electronic properties of back-gated 1L$+$ML devices.} (a) Effective conductivity $GL/W$ as a function of back-gate voltage for the 1L, ML, and 1L$+$ML devices in Fig. 1c. The inset shows $I_{\rm sd}-V_{\rm sd}$ characteristics of the 1L$+$ML device at various gate voltages. (b) The gate-voltage-dependent conductivity of the two-terminal 1L, 2L, and 1L$+$2L devices shown in the inset. The scale bar is 5~$\mu$m.
} \label{fig3}
\end{figure}

\section{Discussion\vspace{-.25em}}
The observations suggest a modified charge state at the 1L/ML interface that impacts longitudinal effective conductivity in the TMDC devices. Although the measurements clearly demonstrate increased effective conductivity and effective field-effect mobility for the composite 1L$+$ML devices, there are several possible interpretations of the underlying mechanism which are not yet clearly distinguished. One attractive explanation is the band bending charge accumulation model as depicted in Fig. 4. When the 1L and ML MoS$_2$ are in contact, electrons from the 1L region diffuse into the ML region because of the potential difference~\cite{workfunction}, resulting in bent bands near the interface, in analogy to the 2DEG in a Al$_{x}$Ga$_{1-x}$As/GaAs heterojunction~\cite{2DEG}. The accumulated electrons are localized at the interface and perhaps experience reduced scattering in this localized state with increased carrier density, thereby promoting conductance in the device. Such a band diagram was also suggested by finite element device simulation~\cite{H} and Kelvin probe force microscopy ~\cite{KelvinP}. The detailed band diagram of a 1L$+$ML interface can be more complex than the traditional band bending depicted in Fig. 4; it has been proposed that the edge state of the ML MoS$_2$ can disrupt the band alignment because of its proximity to the 1L/ML interface~\cite{e}. Nevertheless, this model suggests a mechanism based on established semiconductor heterostructure physics for the accumulation of charge at the interface.

Another possible explanation is the proposed existence of a metallic edge state at the boundary of 2D MoS$_2$ flakes~\cite{edge17, Li2008, Xiao2015, STM}. Considering that the edges of 1L and ML devices are also contacted with Au contacts, any such metallic edge state conductance should also contribute to $\sigma_1$ and $\sigma_{\rm m}$. The conductivity contribution from the edge state, which can be roughly compared using the number of edges divided by the width, should be smaller in 1L$+$ML device ($\sim$3/9~$\mu$m for the device in Fig. 1c) and larger in 1L ($\sim$2/3~$\mu$m, Fig. 1c) or ML ($\sim$2/2.5~$\mu$m, Fig. 1c) device. There is still no direct transport report proving the metallic conductivity at the independent edge of 1L or ML MoS$_2$, and the enhanced conduction that we observe are specific to the 1L/ML interface, not MoS$_2$ edge. Thus the edge state of the 2D material is not likely to be solely responsible for the enhanced effective conductivity and mobility in our measurements.

More relevant here, conductive edge states at MoS$_2$ layer dislocations have also been measured using microwave impedance microscopy~\cite{Wu2016}. Our results of interface band bending and enhanced longitudinal conduction support this picture of a narrow confined low-dimensional conducting charge accumulation, perhaps induced by the band bending known to occur at the 1L/ML interface from the photocurrent measurements.  With now several reports of conducting boundaries between MoS$_2$ regions of different  layer number, additional theoretical and experimental investigation is required to construct a clear interpretation of the origin of these interfacial electronic states and their enhanced conductivity.

\begin{figure}
\centering
\includegraphics[width=10cm]{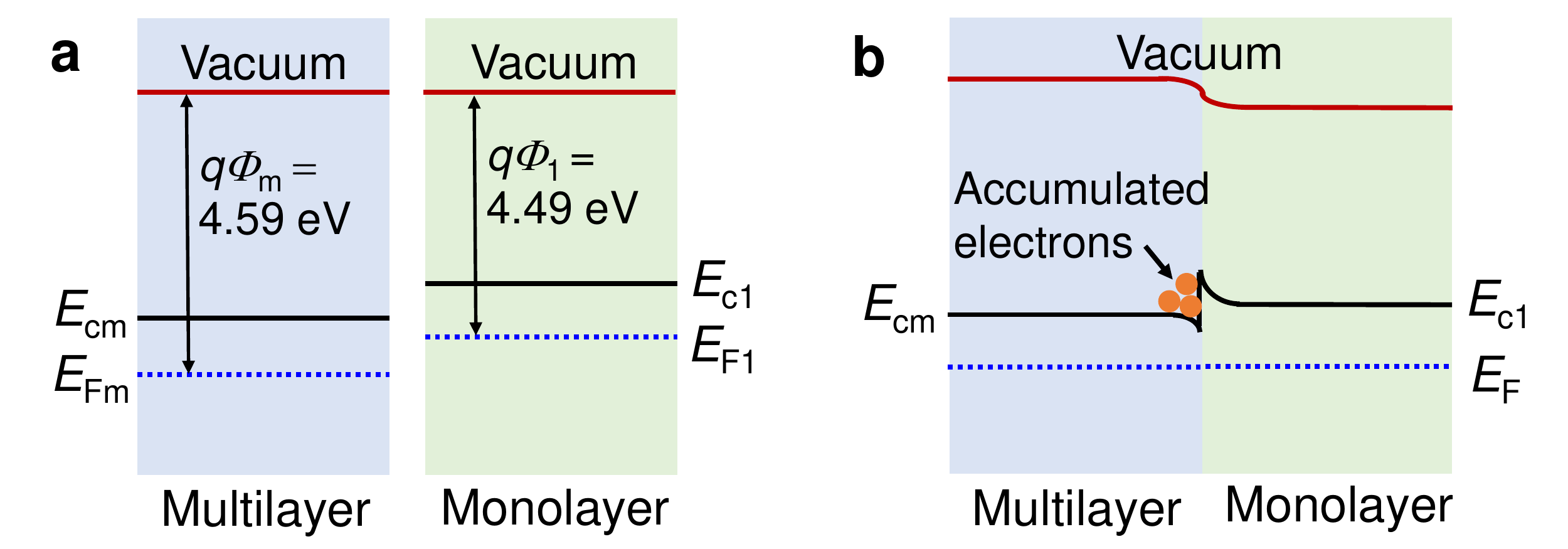}
\caption{{Schematic band diagrams.} (a) Schematic band diagrams of a multilayer and a monolayer MoS$_2$. $q\Phi_{1}$ and $q\Phi_{\rm m}$ are the work functions of 1L and ML MoS$_2$, respectively \cite{workfunction}. (b) Schematic band diagram of MoS$_2$ 1L/ML homojunction in thermal equilibrium.
} \label{figure4}
\end{figure}

Despite the open question of the precise microscopic origin of the enhanced effective conductivity at 1L/ML interfaces, we use a simple electron affinity model to interpret the transport results based on the success of similar band engineering of interface states in 2D interfaces of 3D heterostructures~\cite{semiconductorBook1}. A heterojunction band model is applied to explore the properties of the confined interfacial edge states (See the Supplementary Information). The width of depletion region $x_1$, the width of accumulation region $x_{\rm m}$ and the carrier density at the interface $N_{\rm di}$ can be extracted from the model. The values for the 1L$+$ML device shown in Fig. 1c are $x_1 = 1.2$~nm, $x_{\rm m}= 2.3$~nm, and $N_{\rm di}=7.7\times10^{12}$~cm$^{-2}$. Because of the weaker electron screening in a 2D material with respect to that in a 3D material, the interface electrons can spread over a wider range~\cite{JAP}. Our calculation may underestimate the $x_{\rm m}$ and $x_{\rm 1}$ values. Nevertheless, the value of the depletion width $x_1$ is in reasonable agreement with the band profile at the MoS$_2$/graphene interface (depletion width $\sim 5$~nm)~\cite{STM} directly imaged by scanning tunneling microscopy. $N_{\rm di}$ is about 10 times $N_{\rm dm}$ and $N_{\rm d1}$, suggesting electron accumulation at the interface. Considering that the interface electrons are spatially separated from impurities, a significant scattering source in the interior, they may possess higher mobility than that of interior electrons. Based on the band bending and metallic edge state evidence here and elsewhere~\cite{e, Wu2016}, a highly conductive electronic accumulation is a reasonable explanation for the enhanced longitudinal conduction at 1L/ML interfaces compared to separate 1L and ML devices.

\section{Conclusion\vspace{-.25em}}
In summary, we have fabricated and investigated MoS$_2$ monolayer/multilayer homojunctions with source-drain bias applied longitudinally along the interface. Scanning photocurrent microscopy reveals enhanced photoresponse at the 1L/ML interface, which can be explained by the band bending at the interface. Electronic transport measurements of homojunctions compared to distinct 1L and ML devices provide evidence of enhanced conductivity along the boundary, indicative of a modified interfacial charge state. Although the precise microscopic mechanism driving this enhanced conduction is still not clear, our measurements reveal that the TMDC homojuction interface has a non-trivial impact on longitudinal conductivity of a composite layered device. Further exploration and exploitation of these band engineering edge features in layered heterojunction devices can open a potential pathway to achieve a confined metallic 1D electronic state in TMDCs~\cite{1DSTO}.

\section*{Acknowledgments\vspace{-.25em}}

The work was supported by the Institute for Sustainability and Energy at Northwestern (ISEN) and by the U.S. Department of Energy, Office of Basic Energy Sciences under award number DE-SC0012130 (scanning photocurrent). Preparation of layered samples for this work was partially supported by the National Science Foundation’s MRSEC program (DMR- 1121262) and made use of its Shared Facilities at the Materials Research Center of Northwestern University.  Characterization made use of the NIFTI facility  in Northwestern University's Atomic and Nanoscale Characterization Experimental Center (NUANCE), which has received support from the MRSEC program (NSF DMR-1121262) at the Materials Research Center, the Nanoscale Science and Engineering Center (NSF EEC-0118025/003), the State of Illinois, and Northwestern University. Portions of device fabrication were performed at Northwestern University Micro/Nano Fabrication Facility (NUFAB), which is partially supported by Soft and Hybrid Nanotechnology Experimental (SHyNE) Resource (NSF NNCI-1542205), the MRSEC program (NSF DMR-1121262), the State of Illinois, and Northwestern University. The authors acknowledge V. Chandrasekhar for assistance with facilities and instrumentation. N.P.S. acknowledges support as a Research Fellow of the Alfred P. Sloan Foundation.


\end{document}